\documentclass{optica-article}

\journal{opticajournal}

\articletype{Research Article}

\usepackage{lineno}
\usepackage{soul}
\nolinenumbers

\begin{document}

\title{Emulating the Deutsch-Josza algorithm with an inverse-designed terahertz gradient-index lens}

\author{Ashley N. Blackwell,\authormark{1} Riad Yahiaoui,\authormark{1} Yi-Huan Chen,\authormark{1} Pai-Yen Chen,\authormark{1} Thomas A. Searles,\authormark{1} and Zizwe A. Chase\authormark{1,*}}

\address{\authormark{1}Electrical \& Computer Engineering, University of Illinois Chicago, Chicago, IL 60607, USA\\}

\email{\authormark{*}chase8@uic.edu} 

\begin{abstract}
Photonic systems utilized as components for optical computing promise the potential of enhanced computational ability over current computer architectures. Here, an all-dielectric photonic metastructure is investigated for application as a quantum algorithm emulator (QAE) in the terahertz frequency regime; specifically, we show implementation of the Deustsh-Josza algorithm. The design for the QAE consists of a gradient-index (GRIN) lens as the Fourier transform subblock and silicon as the oracle subblock.  First, we detail optimization of the metastructure through numerical analysis. Then, we employed  inverse design through a machine learning approach to further optimize the structural geometry.  In particular, we improved the lens thickness, in order to enhance the resulting output signal for both balanced and constant functions. We show that by optimizing the thickness of the gradient-index lens through ML, we enhance the interaction of the incident light with the metamaterial leading to a stronger focus of the outgoing wave resulting in more accurate implementation of the desired quantum algorithm in the terahertz.
\end{abstract}

\section{INTRODUCTION} 
Improvement upon  current computing architecture is required in order to enhance computation power, time and resources for industries such as medicine, catalysis, and finance. Traditional digital computers rely heavily on many transistors to function with Moore’s law indicating that the number of transistors on a chip needs to double every two years to increase processing capabilities, power, and efficiency \cite{Liu2021PromisesAP,1677462}. Optical computers present a way to scale down the size of transistors while offering the capabilities of parallel processing, high speed, lower power consumption, and operation at multiple frequencies \cite{Arsenault2004TowardsTS, article} thus exploiting a photonic system can make these devices a prime candidate to realize this possibility \cite{doi:10.1126/science.1142892}, especially in the least exploited THz regime. 

In order to achieve an integrated analog optical computing system, it is necessary to focus the incident wave to the next operational computing component which can be achieved by the use of computational metamaterials. Numerous systems, designed for free space, have been proposed but have been difficult to integrate and bulky in design \cite{wei2022metasurface,Yu2011LightPW, He2022, Xiang2022LSA}. One such metamaterial known as the gradient index (GRIN) lens, can alleviate these issues through being an on-chip device. A prototypical GRIN lens consists of a series of micro-layered structures with holes of different geometries to vary the index of refraction \cite{Stellman2007DesignOG, headland2013beam} that manipulate the propagation of electromagnetic waves \cite{Bai2019NearinfraredTM}. Previous works of GRIN lens metamaterial structures have theoretically and experimentally demonstrated to operate in the THz regime \cite{gaufillet2016dielectric, HernandezSerrano2018ArtificialDS, yang2014efficient} with strong focusing capabilities \cite{paul2010gradient, volk2013plane, neu2010metamaterial} and tunability \cite{zeng2014electrically, maasch2014voltage, moharrami2020tunable}.

Recently, GRIN lenses have been integrated as components of larger photonic devices, consisting of a multitude of subblocks, to realize a new optical computing technology, the  quantum algorithm emulator (QAE). QAEs simulate quantum search algorithms with classical waves via the superposition principle, interference phenomena, and in some cases entanglement which is integral to rapid searching and solving difficult problems that would be time and power consuming on digital computers. QAEs have only been explored in the microwave region where the measured electric field amplitude is the probability amplitude of the equivalent quantum state  \cite{zhang2018implementing}. Zhang et al. proposed a dielectric device, made of Veroclear810, consisting of an oracle subblock, two Fourier transform (FT) subblocks, and a phase plate subblock \cite{zhang2018implementing, wei2022metasurface}. The oracle subblock imprints a spatially phase dependent profile on the incoming wave while the FT subblock and phase plate subblock converts the phase difference for the oracle subblock to amplitude information \cite{zhang2018implementing}. Through this device they were able to simulate Grover’s Algorithm and show that the number of iterations performed on the device was consistent with the efficiency of the quantum search algorithm. 

Metamaterials have reached a high degree of maturity and have recently emerged as a new approach for quantum computing. In this context, Wei et al. proposed a quantum searcher of an on-chip silicon device that consisted of four metasurfaces: an oracle metasurface, two metalenses, and a middle metasurface where different spatial positions of the incident wave showed repeatability in the distribution of the output wave intensity \cite{wei2022metasurface}. Cheng et al. experimentally verified a Deutsch-Jozsa (DJ) algorithm with a millimeter scaled all-dielectric device \cite{cheng2020simulate}.

Here, we report a new design for a quantum algorithm emulator in the THz frequency regime based on a simple platform and optimized by machine learning. The investigated device is composed of a microstructured oracle subblock made of silicon substrate and a FT subblock made of Kapton polyimide. The Kapton film  chosen  acts as a 2D photonic crystal (PhC) slab and proven to exhibit strong electric field confinement and interaction with THz waves with minimal absorption loss \cite{Kyaw2020GuidedmodeRI}. In this work, numerical simulations enhanced by machine learning (ML) were applied to optimize the hole radii and thickness of the FT subblock to achieve an optimal distribution of wave intensities. The structural design of the QAE is first presented along with the numerical analysis showing the initial optimized output. Lastly, the process and results for the ML process are discussed.  Our aim is that the simple design of our device and its compact size may strongly relax the constraints for high frequency domains (e.g., IR and visible) which are expected to play a larger role in photonics and quantum technology as a whole.

\section{NUMERICAL EVALUATION OF QAE}
 
The block diagram of the DJ algorithm is shown in Fig.~1 (top panel). The whole device consists of two functional subblocks, oracle and Fourier transform, respectively, brought into optical contact. Fig. 1 (bottom panel) represents the schematic view of the metamaterial-based quantum emulator. The oracle block is made from a 500-{\textmu}m-thick silicon (Si) substrate. It modulates the electric field profile of the incident THz light by assigning a phase shift of 0 or $\pi$ on each spatial position along $y$-axis. This phase shift is introduced by physically varying the radius of the holes array drilled in the silicon substrate, or oracle block, and by consequence also varies the effective electric permittivity $\epsilon_e$ along the y-axis thereby encoding the function $f(y)$.  

\begin{figure}[htbp]
    \centering
   \includegraphics[scale=.4]{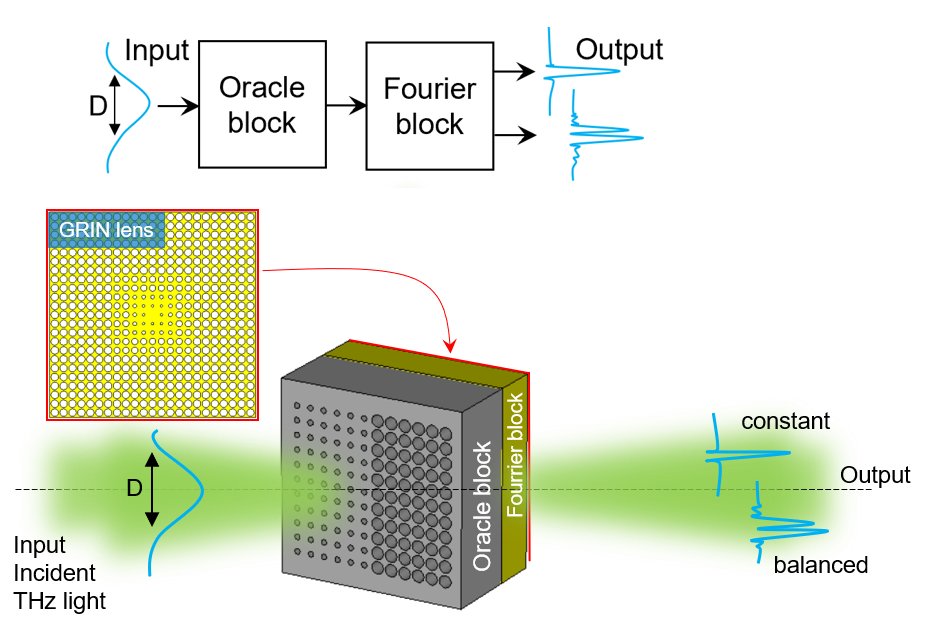}
    \caption{ Schematic of the Quantum Algorithm Emulator. (Top panel) Block diagram of the DJ algorithm and (Bottom panel) metamaterial-based multi-layer configuration of the DJ algorithm emulator. The input THz wave of width D is modulated by the oracle block of varying hole diameters and then transformed through the Fourier block to show the output signal of either a constant function (top output) or balanced function (bottom output).}
    \label{fig1}
\end{figure}

\begin{figure}[htbp]
\centerline{\includegraphics[width=0.7\linewidth]{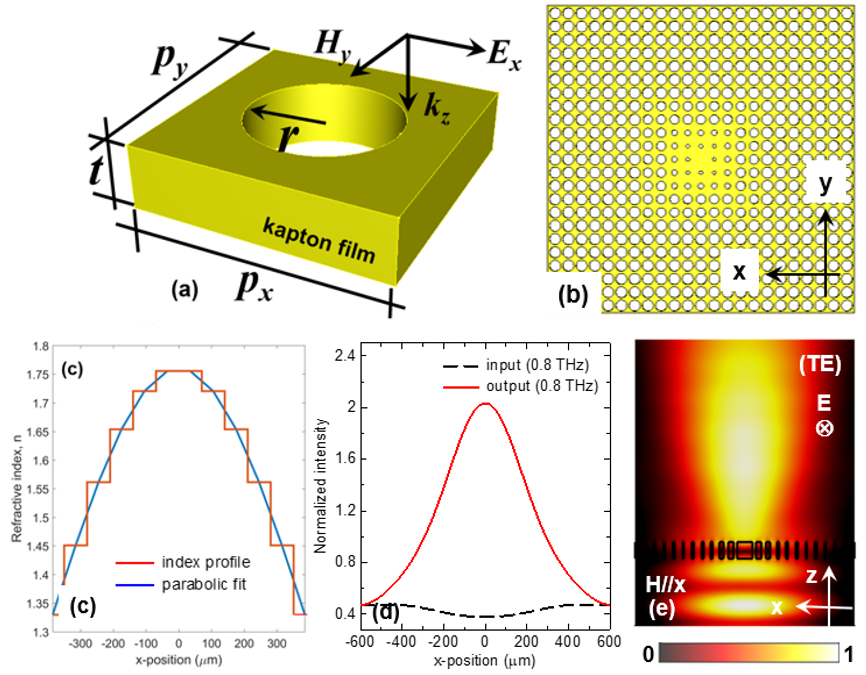}}
\caption{\small{(a) Structural design of the unit cell with the relevant geometrical dimensions: $p_{x}$ = $p_{y}$ = {70 {\textmu}m}, $t$ = {127 {\textmu}m} and $r$ varies 10, 15, 20, 25 and 30 {\textmu}m}, respectively. (b) Schematic view of the GRIN lens. The aperture size of the lens is $\sim$ 1 mm. (c)  Index profile of the structure at 
0.8 THz. The blue solid line is a parabolic fit to the refractive index profile. (d) Simulated electric field intensity on the focal plane at 0.8 THz. The input electric field intensity is also plotted for comparison. (e) Simulated normalized electric field  distribution of the GRIN lens at 0.8 THz, for linearly TE-polarized radiation.}
\label{fig2}
\end{figure}

The Fourier transform block is made from 127-{\textmu}m-thick polyimide film with various hole sizes acting as an all-dielectric  GRIN metalens as shown in Figs.~2(a) and (b).  Each hole has a period of 70 {\textmu}m with varying radii of 10, 15, 20, 25, and 30 {\textmu}m, respectively from the center to the edge of the device. Due to the rotational symmetry imposed by the geometry of the design, the structure has an identical response for both linearly TE-polarized and TM-polarized waves. Also, note that the use of flexible substrates provides an unprecedented route to achieve frequency tunability due to modifications in the profile and the periodicity of the structures when the substrates are manipulated mechanically \cite{Kyaw2020GuidedmodeRI}.

The polyimide film is treated as a dielectric with $\varepsilon = 3.3 + i0.05$. For the GRIN lens design, we chose a radially symmetric refractive index gradient following a parabolic index profile $n(r)= n_0\text{sech}(\alpha{r})$ with  $\alpha=1/r_0 + \cosh^{-1} (n_0/n_r0)$ [Fig. 2(c)]. For this purpose, we introduced a spatial variation of the refractive index by arraying unit cells of different radius such that the refractive index gradually decreased from the center of the GRIN lens. The dimensions of each cell are modified to fit the refractive index profile on the device. The design of the final GRIN lens structure is shown in Fig. 2(b). The effective permittivity of the oracle block can be expressed as: $\epsilon_e(y)=(3\lambda_0/d_0)^{2}$; $\Delta \phi=0$, $\epsilon_e(y)=(2.5\lambda_0/d_0)^{2}$; $\Delta \phi=\pi$, where $d_0$ is the length of the oracle block and $\lambda_0$ the working wavelength.

To examine the performance of the designed metalens, we plotted in Fig.~2(d) the cross-section of the normalized intensity profiles in the plane $y$ = 0 (axial $x$–$z$ plane) around the focal point of the metalens at 0.8 THz. The input electric field intensity is also plotted for comparison. Shown in Fig.~2(e) we present the normalized local electric field distribution of the GRIN lens at 0.8 THz, for linearly TE-polarized radiation. In it, we aim to clearly demonstrate the successful realization of the focusing property of our design. When the incident light irradiates the surface of the oracle block, the phase of the transmitted wave is modulated with a factor {$k_{0}n(y)d_{0}$}, where {$k_{0}=2\pi/\lambda_{0}$} is the vacuum wave vector, {$d_{0}$} is the thickness of the oracle block and {$n(y)$} is the effective refractive index at position {$(y)$}. The detecting function {$f(y)$} is encoded into the input states by assigning a phase modulation on each spatial position {$(y)$} along the input transversal direction. The refractive index of the oracle operator is designed to achieve either 0 or {$\pi$} phase distribution depending on the value of the function {$f$}.  The Fourier transform operator is used to evaluate the final results on the output signal. There are optical processes that can produce Fourier transform of field distribution, such as diffraction and Optical spatial filtering, respectively. Generally, the far field diffraction pattern is observed at infinity. By placing a lens after the diffracting aperture, the plane at infinity is imaged onto the focal plane of the lens. This explains why a lens can perform a Fourier transform.

To evaluate the performance of the device, we performed numerical simulations based on the finite-difference time-domain (FDTD) method. The length scale of the mesh was set to be less than or equal to ${\lambda_{0}}/{10}$ throughout the simulation domain, where ${\lambda_{0}}$ is the central wavelength of the incident radiation. The input and output ports are located at about 10${\lambda_{0}}$ from the device with open boundary conditions.  The blue line plots in Figure 3 show the electric field intensity at the focal plane of the GRIN lens with the detecting function being constant and balance, respectively, computed for the initial numerical analysis at 0.8 Thz. 
The maximum intensity at y = 0 position means the encoded function {$f(y)$} is constant. However, if the center intensity is zero this indicates that the oracle subblock carries a balance function. Since the 0 and $\pi$ phase elements correspond with different effective permittivities, it makes the oracle subblock processes spatially varying impedance which results in different transmissions creating a non-symmetric intensity for the output of the electric field as shown in Fig 3 (blue line, right panel).

\begin{figure}[htbp]
\centering
\includegraphics[width=0.7\linewidth]{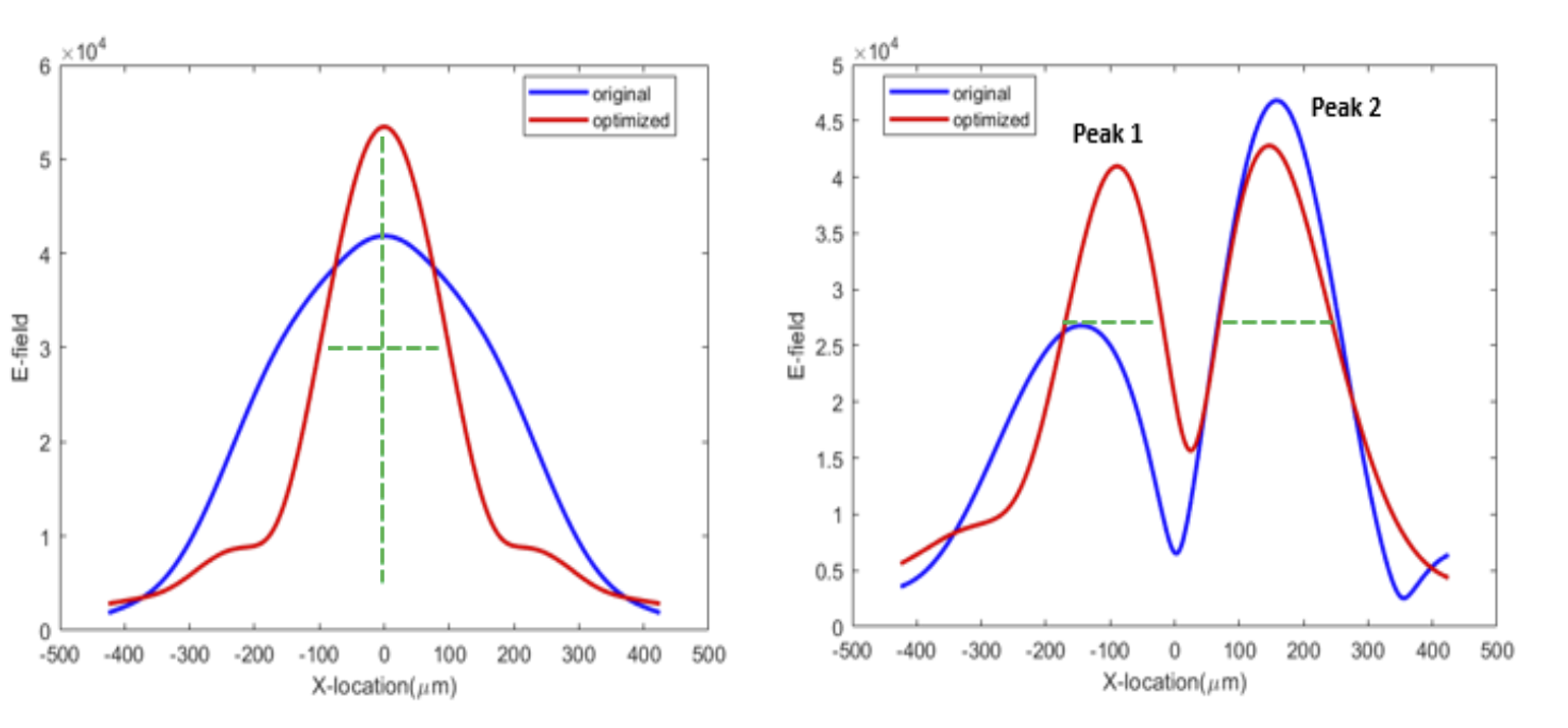}
\caption{\small{The electric field intensity at the focal plane of the GRIN lens for an intermediate working frequency of 0.8 THz, with the detecting function being constant  (left panel) and balance (right panel), respectively. The blue and red lines are simulations of the output pre- and post-application of machine learning based inverse design.}}
\label{fig3}
\end{figure}


\section{MACHINE LEARNING OPTIMIZATION OF QAE}

To further engineer the structural parameters, a machine learning (ML-) based optimization procedure is implemented into the design of the grin lens. ML is a powerful tool for optimization and is expected to be an efficient complementary of electromagnetic wave numerical simulations \cite{ma2019probabilistic, huang2021inverse, zhou2022metamaterials}. Recently, one of the most popular ML techniques used in conjunction with EM field wave simulations is the inverse-design model. Comparing to conventional methods, inverse-design traces the geometry configurations from the output performance such as resonant frequencies or S-parameters\cite{ma2021deep,liu2018generative}. The overall procedure used in this study is shown in Fig. 4. First, the data for training the proposed neural network (NN) model is generated using the numerical simulation software (i.e., CST) based on the initial structure parameters as in Fig. 4(a). 

After which, the geometry configuration is randomly generated using MATLAB with the constraints on $r_{i}$ and h, respectively. The constraints for $r_{i}$ are $0<r_i<a/2$ and $r_{1}<r_{2}<r_{3}<r_{4}<r_{5}<r_{6}$ where $a = 70~\mu m$ is the characteristic length of the unit cell of each hole in Fig.~4(b). Additionaly, the constraint of $h$ is $50~\mu m<h<300~\mu m$. To optimize the performance of the GRIN lens, we divided the holes on the surface into six groups from inner to outer ring with the radius of $r_{1}$, $r_{2}$, $r_{3}$, $r_{4}$, $r_{5}$, and $r_{6}$, respectively. In addition, the thickness of the grin lens h is also involved into the optimized procedure since the path of the wave’s propagation can also affect the performance. Additional details are provided in Appendix B. 

\begin{figure*}[htbp]
\centerline{\includegraphics[width=0.95\linewidth]{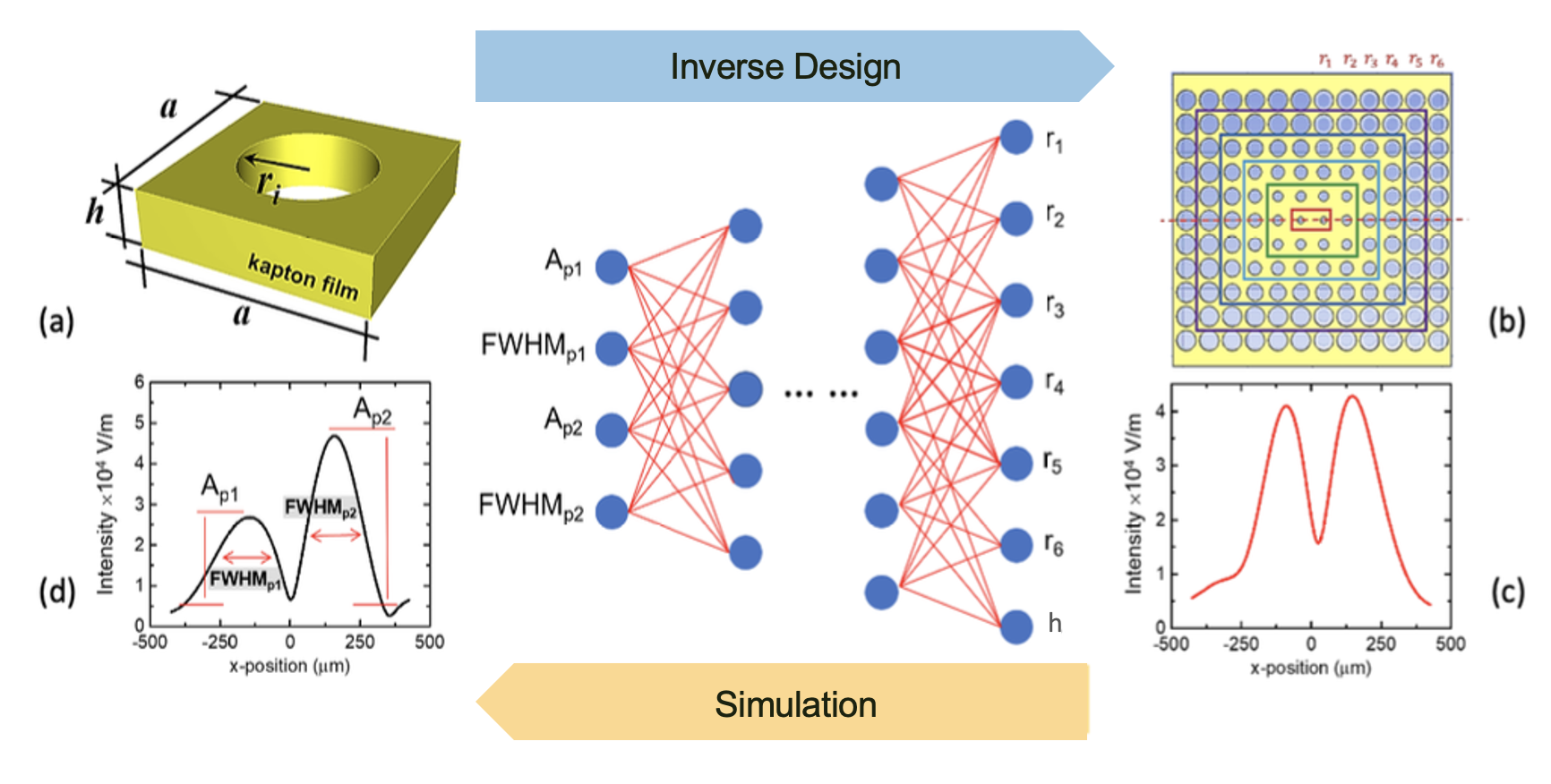}}
\caption{\small{The schematic of the proposed ML-based inverse-design of the Fourier subblock. (a) shows a unit cell which we simulated to generate the whole GRIN lens shown in (b) of varying hole diameters. (c) is the result of the balanced function which is used to apply inverse design to best fit the peaks (d) comparing the initial data of amplitude at $\pi$ and the full width at half maximum (FWHM) of $\pi$ the two peaks as the input parameters.}}
\label{fig4}
\end{figure*}

 According to the Deutsch-Jozsa algorithm, the key point of the design is in fact the performance of the balanced function case since the output of all zero or all one case will absolutely be one peak. It is important that the proposed structure has an ability to distinguish the input status (i.e., constant or balance) from the output performance. Therefore, the proposed optimization procedure mainly focuses on the output spectrum. Hence, we only need to optimize the performance of the balanced function shown in Fig. 4(c). Moreover, since the output of the half-zero half-one case is expected to have two peaks on two sides, the input for the inverse design model can be simplified as the features of the peaks. Here, we choose the amplitude $A_{pi}$ and the full width at half maximum (FWHM) $FWHM_{pi}$ of the two peaks as the input parameters, i.e., 
\begin{equation} \label{eq1}
(\bar v_{in})=[A_{p1},FWHM_{p1},A_{p2},FWHM_{p2}]
\end{equation}

where $p_{1}$ and $p_{2}$ denote the left and right peak, respectively in Fig. 4(d).

The data is then split into the training and validating set with the ratio of $80\%-20\%$. The training set is used to train the proposed NN model, while the validating set is used to check the performance of the model after each training round. Since the length of the input and output vectors are four and seven only, the artificial neural network (ANN) is well enough to achieve the optimal design. There are seven hidden layers in the proposed ANN model, each hidden layer has 16, 16, 32, 32, 16, 16, and 7 channels, respectively. All the layers except for the last one come with a rectified linear activation function (ReLU). The channel in the last layer is the output of the NN model representing the geometry configuration from Fig.~3 which is represented in the following equation:

\begin{equation} \label{eq2}
(\bar v_{out})=[r_1,r_2,r_3,r_4,r_5,r_6,h]
\end{equation}
where $r_1,r_2,r_3,r_4,r_5,r_6$ denote the radii of the six layers and h is the thickness of the GRIN lens.

After training with 100 iterations, the proposed model can precisely predict the geometry configuration from 
$(\bar v_{out})$. Hence, we can feed the model with the desired $(\bar v_{in})$ so that the desired optimal geometry configuration $(\bar v_{out})$ is achieved. However, if the values of $(\bar v_{in})$ exceed the performance limitation of the proposed GRIN lens, $(\bar v_{out})$ may not get the same performance as $(\bar v_{in})$. As a result, it is required to re-validate the connection between $(\bar v_{in})$ and $(\bar v_{out})$ via the numerical simulation software. After several trials, an optimal geometry configuration is achieved and validated by FDTD method. According to Fig.~4, the performance with respect to the output spectral balance is  much better for the inverse-design than the initial design. The detail comparison of the initial and optimal design is represented in Table 1.

\begin{table}[htbp]
\centerline{\includegraphics[scale=0.8]{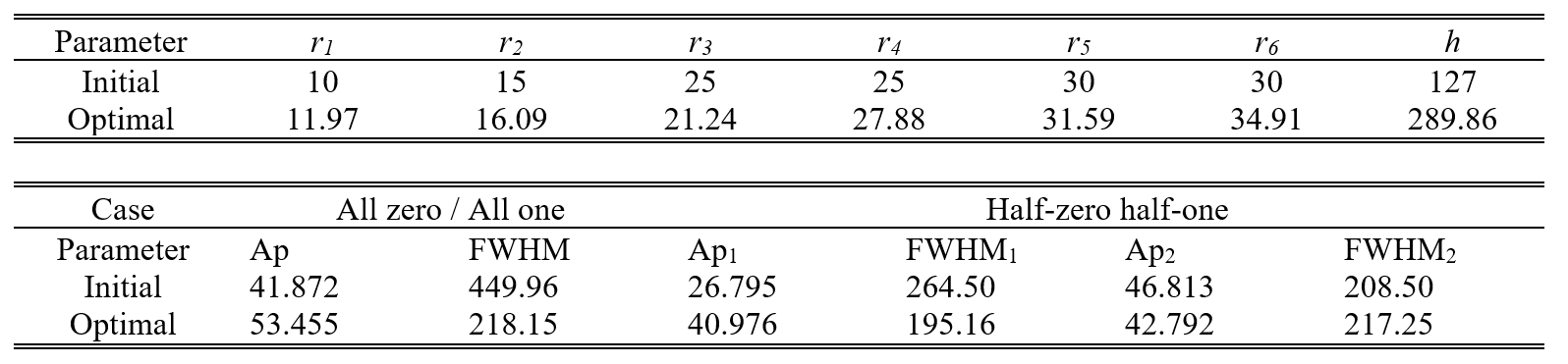}}
\caption{\small{ Top: Comparison of initial numerical GRIN lens parameters to the machine Learning optimized parameters. Bottom: The performance comparison of the initial and optimal design for (i) all-zero all-one case and (ii) half-zero half one case as represented by the output wave characteristics. }}
\label{fig9}
\end{table}

Note that the proposed GRIN lens is based on non-resonating elements that exhibit a weak variation of the refractive index over a wide frequency range, which may strongly relaxes the constraints related to narrow band functioning of traditional metasurfaces. This enables a broadband focusing effect.

From Table 1, the optimized design of the GRIN lens shows minimal increases of all radii except $r_{3}$ with changes being no more than 15\%. However, the biggest change occurred with a factor of two increase of the thickness. This means the radii sizes were near optimal from the initial numerical analysis and that the amount of the computational metamaterial interacting with the incident wave is crucial in achieving the best output amplitudes and full width half maximum (FWHM) for both cases. The thickness of the GRIN lens needs to be at least one-working wavelength in such a way that the wave can better interact with the medium. The thicker the medium the better the interaction will be. However, if the GRIN lens is too thick, the focal point may fall inside the GRIN lens.


For the constant function case the amplitude has been increased by about 26 \% and the FWHM has been decreased by 52\%. For the balanced function, the first peak had its amplitude increased by 50\% and its FWHM decreased by 26\% while the second peak was already close to optimal. For the output of either balanced or constant functions, the sharper peaks (higher amplitude and smaller FWHM) mean greater distribution of the wave intensity, stronger focus of the outgoing wave, and a better probability of handling a higher number of database inputs \cite{zhang2018implementing,wei2022metasurface} as shown in bottom right of Fig. 4 and in the red plot of the right panel of Fig 3. Cheng et al.  also showed an enhancement of their balance function through increasing the number of phase elements in their oracle block.\cite{cheng2020simulate} Our design makes the fabrication process simpler for this improvement in output in that we only have to manufacture a thicker GRIN lens.

\section{Conclusion}
To summarize, we designed, with machine learning, an optimized all-dielectric metadevice as a part of a quantum algorithm emulator for simulating the DJ algorithm in the THz region. Initial structural optimization was constructed using numerical analysis based on FDTD to evaluate its initial performance. Using a ML-based optimization procedure, the original design of the structure was further engineered for an enhanced performance as shown by the two-fold increase in the thickness of the GRIN lens. The resultant optimization  showed performance  improvements in the amplitude and FWHM of the peaks for both the balanced and constant cases and shows promise for the emulation of quantum algorithms with current THz technologies.

\appendix

\renewcommand{\thefigure}{S\arabic{figure}}
\setcounter{figure}{0}

\section{APPENDIX A: DEUTSCH-JOSZA ALGORITHM}
The Deutsch–Jozsa algorithm (Fig. S1) is considered as one of the first examples of quantum algorithms that are exponentially faster than any possible deterministic classical algorithm \cite{deutsch1992rapid}. In the DJ algorithm, the input function is defined as following:

\begin{align}
f(x)={[0,1]}^N  
 \end{align}
 
 The function {$f$} takes n-bit binary values as input and produces either 0 or 1 as output. If the value of {$f$} is 0 on all outputs or 1 on all outputs, {$f$} is called constant function. However, if the value of {$f$} is 1 for half of the output domain and 0 for the other half, {$f$} is called a balance function. The oracle sub-block encodes the detecting function f(y) by assigning a phase shift on each spatial position along the axis of wave propagation while the GRIN lens sub-block acts as a Fourier transformer to evaluate the output signal. At closer inspection for classical light, the incident wave is defined as:
 
\begin{align}
|\psi_{in}\rangle=|0\rangle^{\otimes n}|1\rangle
 \end{align}

A Hadamard transform is applied to this function followed by a unitary transform on the superposition state (Fig. S1)  to produce the following output function of the outgoing wave as represented in the following equation: 
\begin{align}
|\psi_{out}\rangle= [((-1)^{f(0)} + (-1)^{f(1)}) |0\rangle + ((-1)^{f(0)} - (-1)^{f(1)}) |1\rangle ] * (1/\sqrt{2})(|0\rangle -|1\rangle)
 \end{align}
 
As one can see, if f(0) = f(1) there is constructive interference on $|0\rangle$, and destructive interference on the $|1\rangle$ component for the first register. Therefore we obtain a constant function. If the opposite is true, then f(0) not = f(1) and a balanced function is the output \cite{deutsch1992rapid, BJPG2018, Deustsch1985}.
 
\begin{figure}[h]
\centerline{\includegraphics[width=1\linewidth]{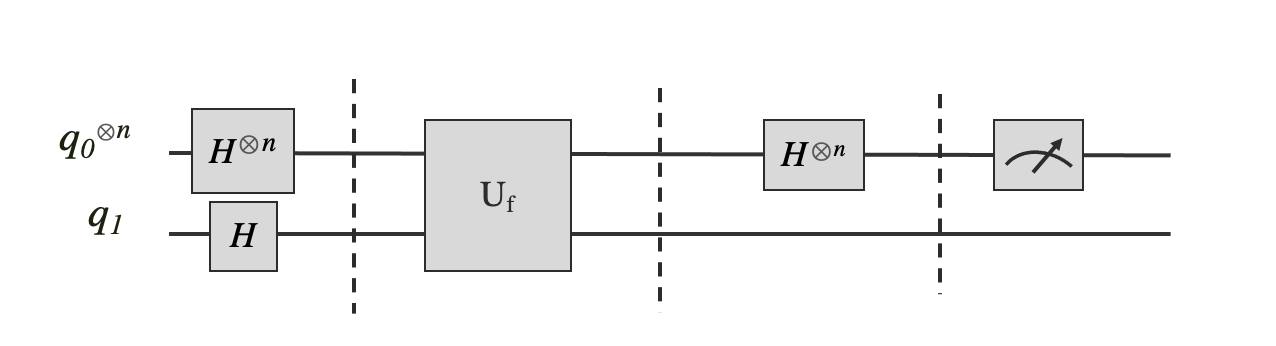}}
\caption{\small{Schematic of the Deutsch-Josza Algorithm}}
\label{fig: Fig. S1}
\end{figure}

\section{APPENDIX B: Additional Machine Learning Optimization Details}

\renewcommand{\thefigure}{S\arabic{figure}}
\setcounter{figure}{1}

The neural network structure contains seven hidden layers with 16, 16, 32, 32, 16, 16, and 7 channels, respectively. All the layers except for the last one come with a rectified linear activation function (ReLU). A softmax function is connected after the last hidden layer to export the output vectors. Smooth L1 loss function and Adam optimizer with learning rate of 0.001 are applied to the proposed model. In addition, the batch size is set as 30 for the small batch training \cite{masters2018}. The amount of the entire dataset is 600, the optimizer is Adam, and the learning rate is 0.001. The model is constructed using Pytorch in Python environment and executed under Ubuntu 16.04 platform with 16 GB RAM, Intel core i7-6700HQ CPU @ 2.6 GHz, and NVIDIA GTX 960M GPU. 

Cost function: Smooth L1 Loss 
\begin{align}
L={[l_1,...,L_N] }^T\\
l(x,y)=mean(L)
 \end{align}

L1 loss: 
\begin{align}
l_n=|x_n-y_n |
 \end{align}

Smooth L1 Loss:
\begin{align}
l_n=(0.5(x_n-y_n )^2)⁄\beta \\
|x_n-y_n |<\beta
\end{align}
or
\begin{align}
l_n=|x_n-y_n |-0.5\beta 
\end{align}

where $\beta$=0.1, $x_n$ and $y_n$ are the input and output elements, respectively. Based on our experiment, the smooth L1 loss can prevent an over-fitting and exploding gradient issue and hence performs better than common loss functions such as mean square error or L1 loss. The training and validation comparison is shown in Fig. S2.  

\begin{figure}[htbp]
\centering
\includegraphics[width=0.75\linewidth]{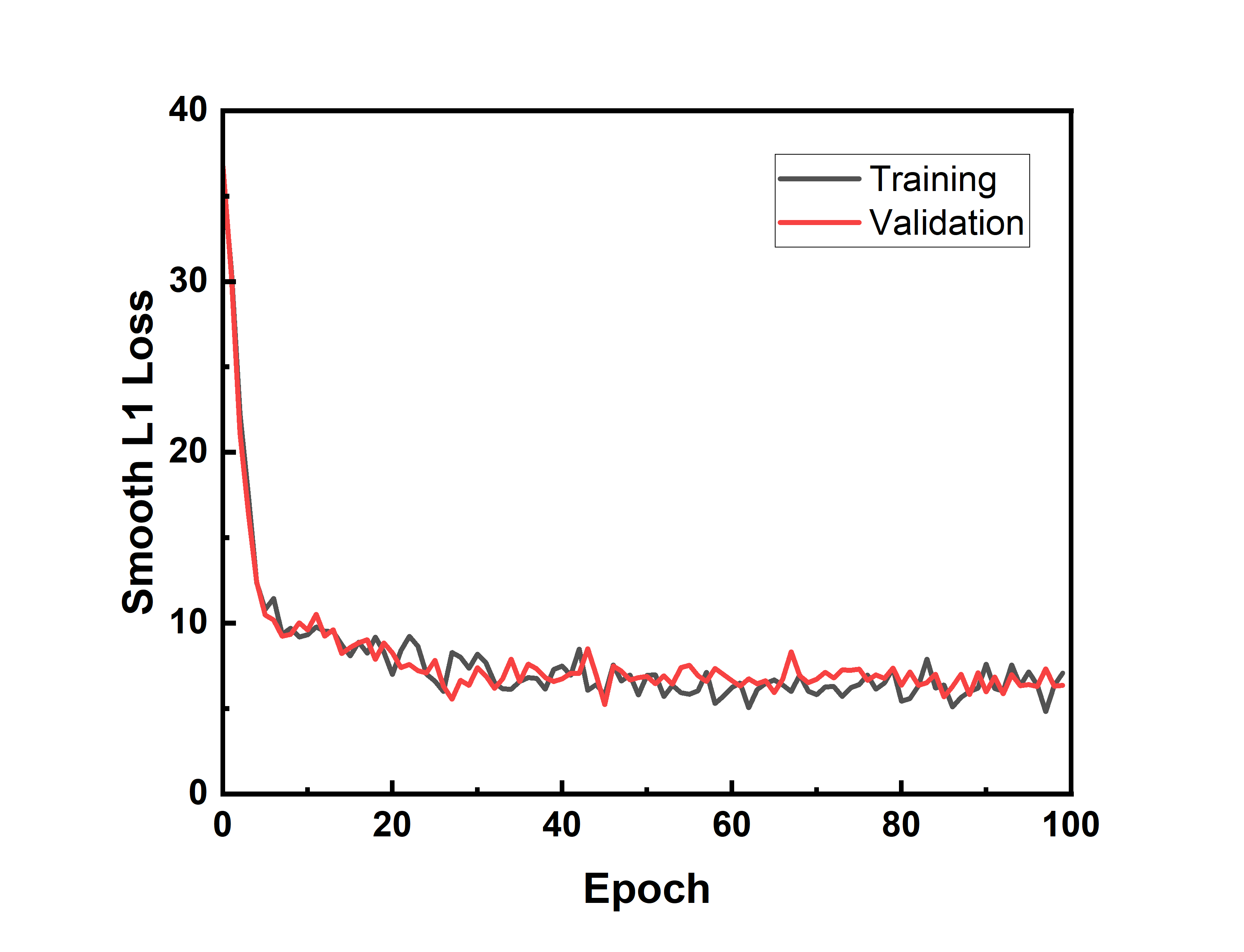}
\caption{\small{The training and validation losses of each epoch. Since the output is still required to import back to the numerical simulation to demonstrate the performance, the loss value is sufficient to examine the performance of the proposed model instead of transferring to percent error.}}

\label{fig:Fig. S2}
\end{figure}

To validate the consistency, we executed the experiment five times and evaluated the coefficient of variation (CV) for each configuration. The result is as shown in the Table S2, which demonstrates that there should not be a one-to-many proposed in the proposed model.

\renewcommand{\thetable}{S\arabic{table}}
\setcounter{figure}{0}

\begin{table}[htbp]
\centerline{\includegraphics[scale=0.45]{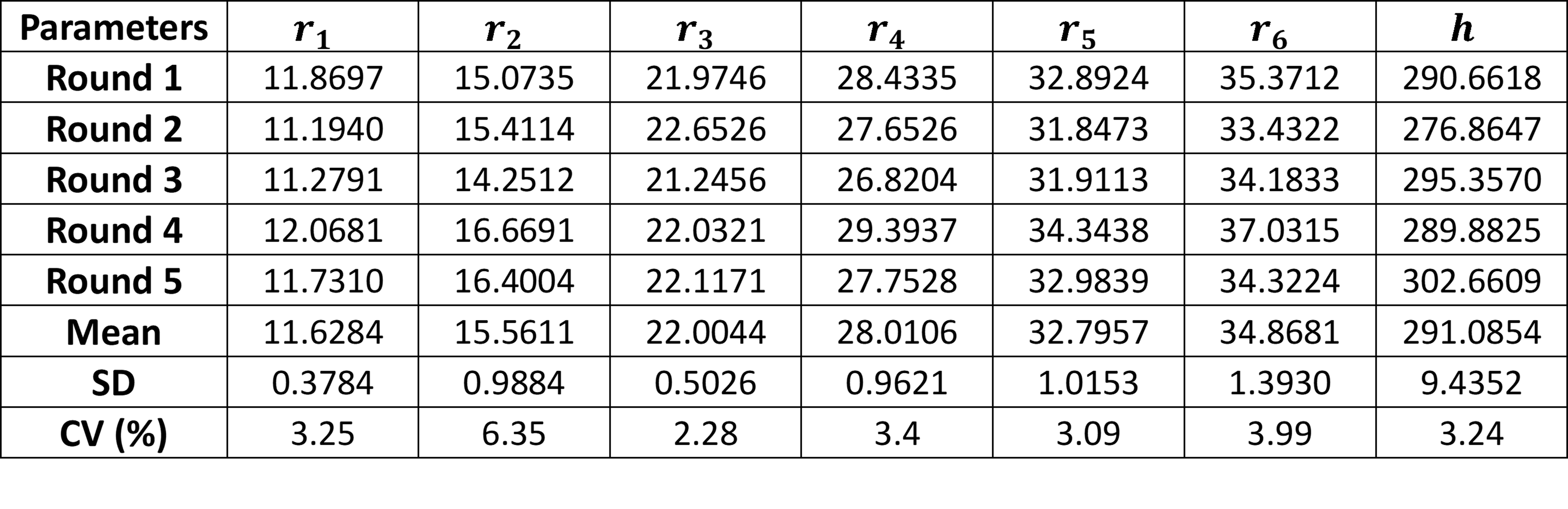}}
\caption{\small{Each round represents a computational evaluation of all radii and the thickness of the Fourier Transform block with statistical data given.}}
\label{table: Table S1}
\end{table}

\newpage
\section*{Acknowledgment}
A. N. B. would like to thank the partial financial support of the GEM Fellowship for this work. Further, Z. A. C. would like to thank the Bridge-to-Faculty Fellowship. 

\section*{FUNDING}
This material is based upon work supported by the U.S. Department of Energy, Office of Science, National Quantum Information Science Research Centers, Co-design Center for Quantum Advantage (C2QA) under contract number DE-SC0012704.

\section*{DISCLOSURES}
N/A

\bibliography{DJAlgo}

\end{document}